\documentclass[useAMS,usenatbib,onecolumn]{mn2e}

\usepackage{graphicx}
\usepackage{amssymb}
\usepackage{amsmath}
\usepackage{times}
\usepackage[]{eufrak}
\usepackage{color}

\newcommand{\ie}{\textit{i.e.},~}

\newcommand{\be}{\begin{equation}}
\newcommand{\ee}{\end{equation}}
\newcommand{\bdm}{\begin{displaymath}}
\newcommand{\edm}{\end{displaymath}}
\newcommand{\bea}{\begin{eqnarray}}
\newcommand{\eea}{\end{eqnarray}}

\title[General Relativistic Magnetospheres of Rotating
Oscillating Neutron Stars]{General Relativistic Magnetospheres of
Slowly Rotating and Oscillating Magnetized Neutron Stars}

\author[V.~S.~Morozova, B.~J.~Ahmedov and O.~Zanotti]
        {Viktoriya~S.~Morozova$^{(1,\;2,\;3)}$, Bobomurat~J.~Ahmedov$^{(1,\;2,\;3)}$ and
        Olindo~Zanotti$^{(4)}$
                                                                \\
 $^{(1)}$Institute of Nuclear Physics,
        Ulughbek, Tashkent 100214, Uzbekistan                   \\
        $^{(2)}$Ulugh Beg Astronomical Institute,
        Astronomicheskaya 33, Tashkent 100052, Uzbekistan       \\
        $^{(3)}$The Abdus Salam International Centre
for Theoretical Physics, 34014 Trieste, Italy\\
$^{(4)}$Max-Planck-Institut f$\ddot{u}$r
        Gravitationsphysik, Albert-Einstein-Institut, 14476
        Golm, Germany\\
\\}
\pagerange{\pageref{firstpage}--\pageref{lastpage}}

\pubyear{2010}

\begin{document}

\maketitle

\label{firstpage}

\begin{abstract}

We study the magnetosphere of a slowly rotating magnetized neutron
star subject to toroidal oscillations in the relativistic regime.
Under the assumption of a zero inclination angle between the
magnetic moment and the angular momentum of the star, we analyze
the Goldreich-Julian charge density and derive a second-order
differential equation for the electrostatic potential. The
analytical solution of this equation in the polar cap region of
the magnetosphere shows the modification induced by stellar
toroidal oscillations on the accelerating electric field and on
the charge density. We also find that, after decomposing the
oscillation velocity in terms of spherical harmonics, the first few
modes with  $m=0,1$ are responsible for energy losses that are
almost linearly dependent on the amplitude of the oscillation and
that, for the mode $(l,m)=(2,1)$, can be a factor $\sim8$ larger
than the rotational energy losses, even for a velocity oscillation
amplitude at the star surface as small as $\eta=0.05 \ \Omega \
R$. The results obtained in this paper clarify the extent to which
stellar oscillations are reflected in the time variation of the
physical properties at the surface of the rotating neutron star,
mainly by showing the existence of a relation between $P\dot{P}$
and the oscillation amplitude. Finally, we propose a qualitative
model for the explanation of the phenomenology of intermittent
pulsars in terms of stellar oscillations that are periodically
excited by star glitches.

\end{abstract}


\section{Introduction}

The theoretical study of radio pulsars dates back to the
work of \cite{Goldreich1969} who first suggested
 the existence of
a magnetosphere with a charge-separated plasma around rotating
magnetized neutron stars. A spinning magnetized neutron star
generates huge potential differences between different parts of
its surface. The cascade generation of electron-positron plasma in
the polar cap region already proposed by \citet{Sturrock1971} and
\citet{Ruderman1975} requires that the magnetosphere of a neutron
star is filled with plasma, thus screening the longitudinal
electric field and bringing the plasma into co-rotation with the
neutron star. Because co-rotation is not possible outside the light
cylinder (the radius $R_{LC}=c/\Omega$ at which the co-rotation
speed equals the speed of light), essentially two different
topologies of the magnetic field lines are naturally produced:
closed lines, namely those returning to the stellar surface, and
open lines, i.e. those crossing the light cylinder and going to
infinity. As a result, plasma may leave the neutron star along the
open field lines and it is generally thought that pulsar radio
emission is produced in the region of open field lines well inside
the light cylinder and within a given angle from the polar axis.

Beside the seminal papers by   \cite{Goldreich1969},
\cite{Sturrock1971},  \cite{Ruderman1975}, \cite{Mestel1971} and
\cite{Arons1979}, pulsar magnetospheres have been investigated by
a large number of authors over the years. We only mention here the
reviews by \cite{Arons1991}, \cite{Mestel1992} and
\cite{Muslimov1997}, where subsequent achievements and some new
ideas have been presented. Thorough description of known
magnetosphere properties may be found, for example, in the book of
\cite{Beskin2009}. It should also be mentioned that in the last
few years time dependent numerical simulations of neutron star
magnetospheres have been proposed as a new promising tool for
investigating the complex physics of these systems. At least
qualitatively, the numerical approach has confirmed the most
fundamental features of what was expected from the stationary
solution of the Grad-Shavranov equation~\citep{Contopoulos1999,
Gruzinov2005}, such as the existence of closed magnetic field
lines up to the light cylinder~\citep{Komissarov2006,
McKinney2006a}, or the scaling of the spin down luminosity on the
angular velocity and on the inclination angle of the neutron star
angular momentum with respect to its magnetic
moment~\citep{Spitkovsky2006}. In spite of this spectacular
progress, however, the numerical approach still suffers from some
serious limitations, such as the lack of a unified scheme in which
both the force free regime and the plasma regime of
magnetohydrodynamics are simultaneously taken into account, or the
lack of a consistent treatment of resistive effects in the current
sheet.

The analytic approach, on the other hand, can still provide a deep
understanding of pulsar physics. In particular, a lot of attention
has  been paid to the existence of a strong electric field
induced by the rotation of the star, as already noticed by
\cite{Deutsch1955}. More recently,~\cite{Beskin1990} and,
independently, \cite{Muslimov1990} were the first to find that the
frame dragging induced by general relativistic effects provides a
source of additional electric field contributing to particle
acceleration in the polar cap region. The accelerating component
(parallel to the magnetic field) of the  electric field is driven
by deviations of the space density charge from the
Goldreich-Julian (GJ) charge density, which is determined by the
magnetic field geometry. As noted by several authors
~\citep{Beskin1990, Muslimov1990, Muslimov1997, Dyks2001,
Mofiz2000, Morozova2008}, the corrections of general relativity in the
plasma magnetosphere of rotating neutron stars are first-order
in the angular velocity of the dragging of inertial frames
and have to be carefully included in any self-consistent model of
pulsar magnetosphere,  especially when computing the resulting
electromagnetic radiation.

Tightly related to this aspect is the possibility that neutron
star oscillations, most likely excited during a glitch phenomenon
(sudden change of the rotational period), propagate into the
magnetosphere, thus affecting the acceleration properties in the
polar cap region. The first attempt to generalize the
Goldreich-Julian formalism to the case of an oscillating neutron
star was made by~\cite{Timokhin2000}, who developed a general
procedure for calculating the GJ charge density in the near zone
of an oscillating neutron star. Using this procedure, the GJ
charge density and the electromagnetic energy losses were computed
for the case of toroidal oscillations at the neutron star surface.
A similar approach has been recently extended to the general
relativistic context by~\cite{Abdikamalov2009} who, just like
~\cite{Timokhin2000}, based their results on the so called low
current density approximation, i.e. on the assumption that the
magnetic field is mainly produced by volume currents inside the
neutron star and by surface currents on its surface, while the
magnetic field due to magnetospheric currents can be neglected. In
the paper of~\cite{Abdikamalov2009} the influence of oscillations
to the magnetosphere electrodynamics was considered for the case
of a non-rotating Schwarzschild star. In the present paper we apply
some of the results of~\cite{Abdikamalov2009} to investigate how
oscillations, produced at the star surface, reflect in the energy
losses from the polar cap region of the magnetosphere of slowly
rotating neutron star. In this respect we extend the work of
\cite{Muslimov1997} by performing a local analysis in the domain
of open magnetic field lines in the inner magnetosphere and taking
into account the effects of toroidal oscillations excited at the
star surface.

The plan of the paper is as follows. In
Sec.~\ref{Goldreich-Julian_relativistic_charge_density} we provide
the minimum general relativistic formalism for understanding
neutron star electrodynamics and we perform a detailed analysis of
the GJ charge density of slowly rotating and oscillating neutron
star. In Sec.~\ref{Poisson_equation} we derive a version of the
Poisson equation that takes into account both general relativistic
effects and the oscillating behavior of the magnetosphere of the
rotating star. Sec.~\ref{Energy_losses}, on the other hand, is
devoted to the computation of the energy losses induced by
oscillations together with rotation. In
Sec.~\ref{Connection_to_the_phenomenology_of_part-time_pulsars} we
propose and motivate a suggestive idea to explain the
phenomenology of intermittent pulsars in terms of the excitation
of stellar oscillations. Finally, Sec.~\ref{concl} contains the
conclusions of our work.

Throughout, we assume a signature $\{-,+,+,+\}$ for the space-time
metric and we use Greek letters (running from $0$ to $3$) for
four-dimensional space-time tensor components, while Latin letters
(running from $1$ to $3$) will be employed for three-dimensional
spatial tensor components. Moreover, we set $c = G = 1$
(however, for those expressions with
an astrophysical application we have written the speed of light
explicitly).

\section{Goldreich-Julian relativistic charge density}
\label{Goldreich-Julian_relativistic_charge_density}

In the slow limit approximation, the spacetime around a
rotating neutron star of total mass $M$, angular momentum $J$ and
angular velocity $\Omega$
is given by \citep{Hartle1968,Landau-Lifshitz2}
\begin{equation}
\label{metricNUT}
ds^2=-N^2dt^2+N^{-2}dr^2+r^2\left(d\theta^2+\sin^2\theta d\phi^2
\right) -2\omega_{\rm{LT}} r^2\sin^2\theta d\phi dt\ ,
\end{equation}
where $N\equiv (1-2M/r)^{1/2}$ is the gravitational lapse
function, while $\omega_{\rm{LT}}=2aM/r^3$ is the Lense-Thirring
angular velocity, which represents the angular velocity of a
freely falling inertial frame. The specific angular momentum $a$,
on the other hand, is defined as $a=J/M$, $J$ is the angular
momentum of the star. We note that the metric \eqref{metricNUT} is
split according to the $3+1$ formalism of general relativity
\citep{Arnowitt62}, which admits a natural Eulerian observer, also
called the ZAMO (zero angular momentum observer), with
four-velocity $n_\alpha$ given by
\begin{equation}
n_\alpha = \left\{-N,0,0,0\right\} \ .
\end{equation}
In the rest of our discussion, when we introduce any
three-vector, like for instance the electric field, we
assume that it is
defined in the locally flat spacetime of the ZAMO
observer, and we denote its orthonormal
components with hat superscripts.

From the system of Maxwell equations and after assuming the
magnetic field of a neutron star to be stationary in the
co-rotating frame, \cite{Muslimov1992} derived
the following Poisson equation for the scalar potential $\Phi$
\begin{equation}
\label{Poiss}
{\mathbf\nabla}\cdot\left(\frac{1}{N}{\mathbf\nabla}\Phi\right)=
-4\pi(\rho-\rho_{\rm{GJ}})\ \ ,
\end{equation}
where
$\rho-\rho_{\rm{GJ}}$ is the effective space charge density
responsible for the generation of an unscreened electric
field parallel to the magnetic field,
while $\rho_{\rm{GJ}}$ is the Goldreich-Julian charge density that
we discuss below.
In their pioneering work, \cite{Goldreich1969} showed that
a strongly magnetized and highly conducting neutron star, rotating
about the magnetic axis, would spontaneously build up a charged
magnetosphere. In a nutshell, the argument is the following:
if a magnetized rotating neutron star is placed in
vacuum,
enormous unbalanced
electric forces parallel to the magnetic field {\bf B}
would set up at the surface of the star, extracting
charges from the surface into the external vacuum
region, thus producing a filled magnetosphere.
Therefore, \cite{Goldreich1969} hypothesized that a far better
approximation for the magnetosphere would be obtained by shorting
out the component of the electric field {\bf{E}} along {\bf{B}}.
The magnetospheric charges that
maintain ${\bf{E}}\cdot{\bf{B}}=0$ are themselves subject to the
${\bf{E}}\times{\bf{B}}$ drift that sets them into co-rotation with
the star~\citep{Mofiz2000}.
A derivation of the Goldreich-Julian charge density in
the presence of oscillations but in the Newtonian
framework has been performed by~\citet{Timokhin2007}.
Here we discuss the corrections to the standard
Goldreich-Julian charge density when both relativistic
effects and stellar oscillations are taken into account.
Our starting point is the general expression for the
Goldreich-Julian charge density that takes
into account the contribution of the electric field
induced by arbitrary stellar oscillations, \ie
\begin{equation}
\label{GJ}
 \rho_{\rm{GJ}}=-\frac{1}{4\pi
c}\nabla\cdot\left[\frac{1}{N}(\vec{u}-\vec{w})\times\vec{B}
+\frac{1}{N}\vec{\delta v}\times\vec{B}\right]=-\frac{1}{4\pi
c}\nabla\cdot\left[\frac{1}{N}\left(1-\frac{\kappa}{\bar{r}
^3}\right)\vec{u}\times\vec{B} +\frac{1}{N}\vec{\delta
v}\times\vec{B}\right]\ ,
\end{equation}
where $\vec{u}-\vec{w}= \left(\Omega - \omega_{\rm{LT}}
\right)r\sin\theta e_{\hat\varphi}$, while $\vec{\delta v}$ is the
oscillation velocity. Moreover, $R$ is the star radius,
$\bar{r} =r/R$ is the dimensionless radial
coordinate,
$\varepsilon=2M/R$ is the compactness parameter,
$\beta=I/I_0$ is the moment of inertia of the star in units of $I_0=MR^2$
and $\kappa=\varepsilon\beta$.

\begin{figure*}
\begin{center}
\includegraphics[width=8.5cm,angle=0]{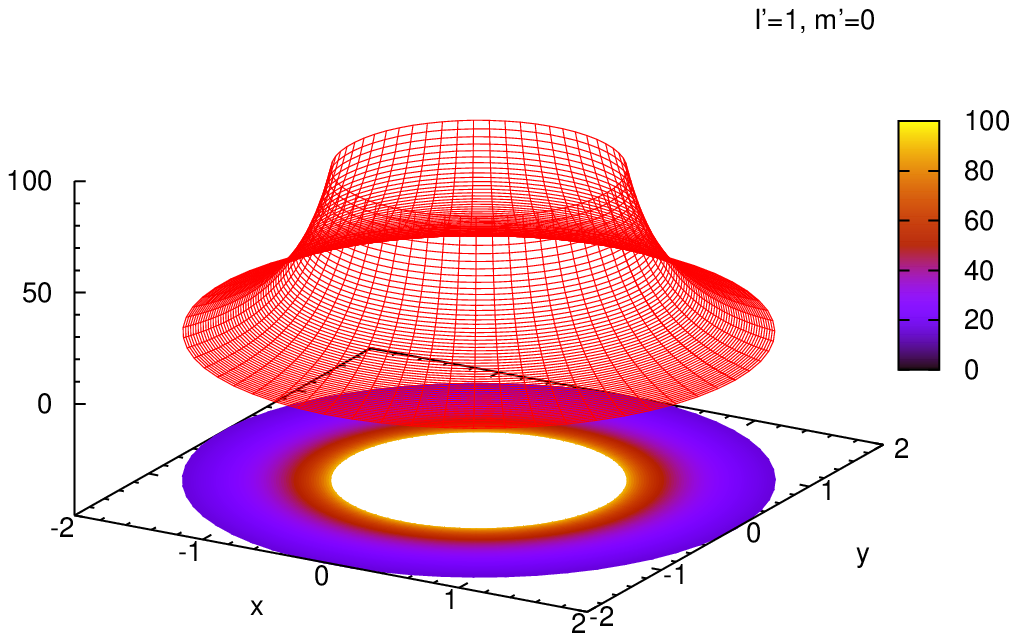}
\hspace{0.125truecm}
\includegraphics[width=8.5cm,angle=0]{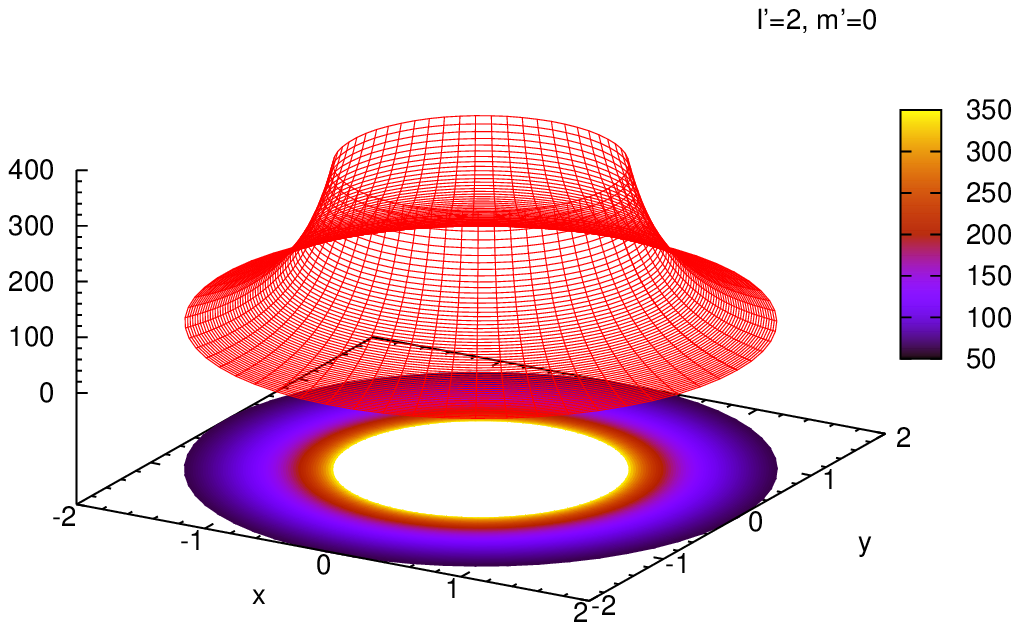}
\includegraphics[width=8.5cm,angle=0]{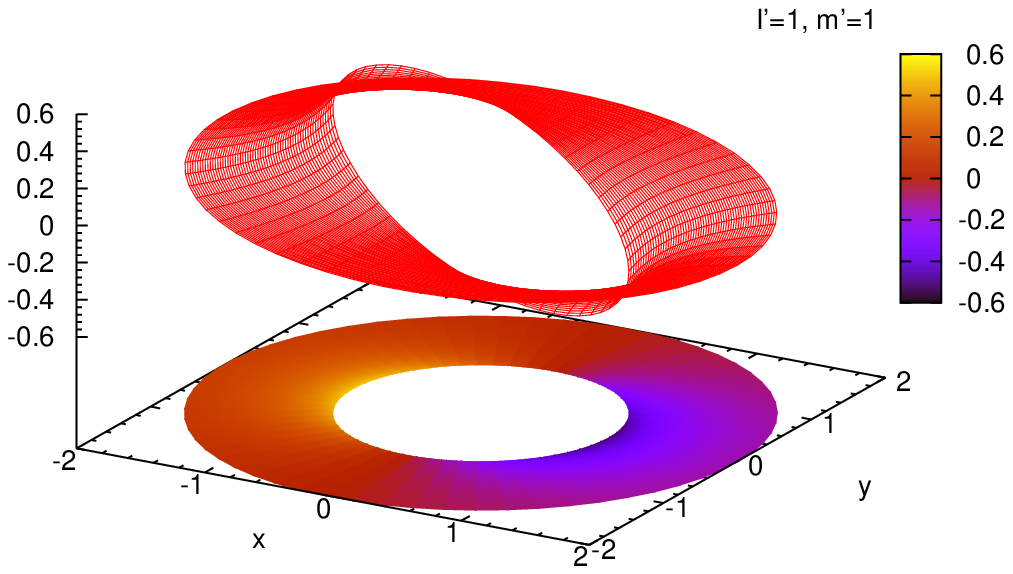}
\hspace{0.125truecm}
\includegraphics[width=8.5cm,angle=0]{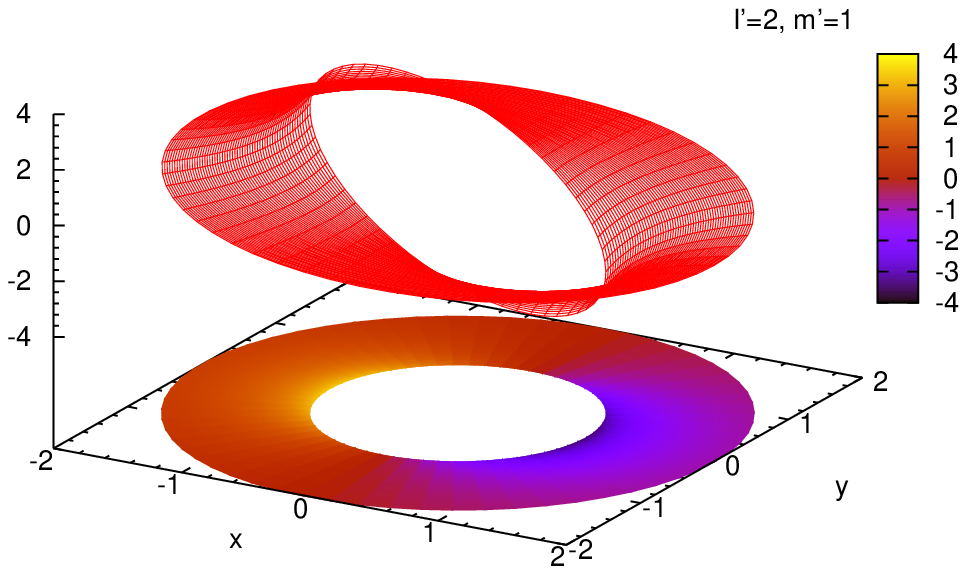}
\end{center}
\vspace*{-0.7cm}
\caption{
Ratio $\delta\rho_{\rm{GJ, l^\prime
      m^\prime}}/\rho_{\rm{GJ},0}$ for the mode $(1,0)$
(left top panel),  for the mode $(1,1)$ (left bottom
  panel), for the mode  $(2,0)$ (right top panel) and for
  the mode $(2,1)$ (right bottom panel).
The representative star parameters are chosen to be $\kappa=0.15$,
$\varepsilon=1/3$, $K=0.01$, $\Theta_0=0.008$, $\Omega=1 {\rm
rad/s}$. } \label{rhoGJ_2D_color}
\end{figure*}

We apply expression (\ref{GJ}) to the case of toroidal
oscillations, whose velocity has components
[see, for example, Eq.~(13.71) of~\citet{Unno1989}
\begin{equation}
\label{vel} \delta v^{\hat{i}}=\left\{0,
\frac{1}{\sin\theta}\partial_{\phi}Y_{l\prime m\prime}(\theta,\phi),
-\partial_{\theta}Y_{l\prime  m^\prime}(\theta,\phi)\right\}\tilde{\eta}(r)e^{-i\omega
t}\ ,
\end{equation}
where $\omega$ is the real part of the oscillation frequency, while
$\tilde{\eta}$ is the radial eigenfunction expressing the
amplitude of the oscillation.
We have
used multipolar indices $\ell^{\prime}$ and $m^{\prime}$ to
distinguish the harmonic dependence of the velocity perturbations
from the harmonic dependence, in terms of $\ell$ and $m$, of the
electromagnetic fields, since these indices are in general
distinct.
As usual, the spherical orthonormal
functions $Y_{lm}(\theta,\phi)$
are the eigenfunctions of the
Laplacian in spherical coordinates. They are given by
\begin{equation}
\label{Y}
Y_{lm}(\theta,\phi)=\frac{1}{\sqrt{2\pi}}e^{im\phi}\Theta_{lm}(\cos\theta)\
,
\end{equation}
where the functions $\Theta_{lm}(\cos\theta)$ satisfy the
differential equation
\begin{equation}
\label{YY}
\frac{1}{\sin\theta}\frac{d}{d\theta}\left(\sin\theta\frac{d\Theta_{lm}(\cos\theta)}{d\theta}\right)
-\frac{m^2}{\sin^2\theta}\Theta_{lm}(\cos\theta)+l(l+1)\Theta_{lm}(\cos\theta)=0
\end{equation}
and can be written as
\begin{equation}
\Theta_{lm}(\cos\theta)=(-1)^m\sqrt{\frac{2l+1}{2}\frac{(l-m)!}{(l+m)!}}P^m_l(\cos\theta)\ ,
\end{equation}
where $P^m_l(\cos\theta)$ are the Legendre polynomials.
For simplicity, we limit our attention to the case in
which the magnetic moment $\mu$ of the star is aligned with its
angular momentum, and, furthermore, we assume that the magnetic
field is a dipolar one, with orthonormal components that
take the form \citep{Muslimov1992, Ginzburg1964}
\begin{equation}
\label{B} B^{\hat r}=B_0\frac{f(\bar{r} )}{f(1)}\bar{r}
^{-3}\cos\theta\ , \ \ B^{\hat\theta} =\frac{1}{2} B_0
N\left[-2\frac{f(\bar{r} )}{f(1)}+ \frac{3}{(1-\varepsilon/\bar{r}
)f(1)}\right]\bar{r} ^{-3}\sin\theta\ ,
\end{equation}
where
\begin{equation}
\label{f} f(\bar{r} )=-3\left(\frac{\bar{r}
}{\varepsilon}\right)^3\left[\ln
\left(1-\frac{\varepsilon}{\bar{r}
}\right)+\frac{\varepsilon}{\bar{r} }
\left(1+\frac{\varepsilon}{2\bar{r} }\right)\right]\ ,
\end{equation}
and $B_0\equiv 2\mu/R^3$ is the Newtonian value of the magnetic
field at the pole of the star.
When the oscillation velocity and the
magnetic field
of the star are given, respectively,
by Eq.~\eqref{vel} and Eq.~\eqref{B},
then Eq.~(\ref{GJ}) provides
the following modified Goldreich-Julian charge density
\begin{eqnarray}
\label{rhoGJ} \rho_{\rm{GJ}}&=&-\frac{\Omega B_0}{2\pi c
}\frac{1}{N\bar{r} ^3}\frac{f(\bar{r}
)}{f(1)}\left(1-\frac{\kappa}{\bar{r} ^3}\right) \nonumber
\\
&-&\frac{1}{4\pi c}\frac{1}{R\bar{r} ^4}B_0 e^{-i\omega
t}\Bigg\{-\frac{1}{N}\frac{f(\bar{r} )}{f(1)}\tilde{\eta}(\bar{r}
)\cot\theta
\Bigg[\frac{\partial}{\partial\theta}\sin\theta\frac{\partial}{\partial\theta}Y_{l^\prime m^\prime}(\theta,\phi)
+\frac{1}{\sin\theta}\frac{\partial^2}{\partial\phi^2}Y_{l^\prime m^\prime}(\theta,\phi)\Bigg]
\nonumber
\\ &+&\frac{N}{2}\left[-2\frac{f(\bar{r} )}{f(1)}+
\frac{3}{(1-\varepsilon/\bar{r}
)f(1)}\right]\sin\theta\frac{\partial}{\partial\theta}Y_{l^\prime m^\prime}(\theta,\phi)
\frac{\partial}{\partial\bar{r} }\frac{\bar{r}
}{N}\tilde{\eta}(\bar{r} )\Bigg\}\ .
\end{eqnarray}
From the point of view of the emission of energy, that we
consider with greater detail in Sec.~\ref{Energy_losses},
the most interesting
region of the magnetosphere is the so called {\it polar cap
region}, i.e. the region where the magnetic field
lines remain open and at distances from the star surface
much smaller than the
light cylinder radius.
If we denote by $\Theta_0$ (computed as explained in
Sec.~\ref{Energy_losses}) the co-latitude of
the last closed magnetic field  line at the star surface,
then the angle $\Theta$ of the last closed magnetic
field line as a
function of $\bar{r} $ is well approximated as
\citep{Muslimov1992}
\begin{equation}
\label{Theta}
\Theta(\bar{r} )\cong\sin^{-1}\left\{\left[\bar{r}
\frac{f(1)}{f(\bar{r} )}\right]^{1/2} \sin\Theta_0\right\}\ .
\end{equation}
It should be remarked that on the surface of the star,
and even within reasonably large
distances far from it, the aperture angles of open magnetic
field lines remain small.
For example, for a neutron star with $\epsilon=0.3$ and
$\Theta_0=0.087\approx 5^{\circ}$, we find
$\Theta=0.197\approx 11.3^{\circ}$ at $r=4 R$.
As a result,
in this approximation the
last term in the curl brackets on the right hand side of
Eq.~(\ref{rhoGJ}) can be neglected, since it contains
$\theta$, with $\theta\leq \Theta(\bar{r} )$, to a second power larger
than other terms in the same brackets.
Taking into account Eq.~(\ref{Y}) and (\ref{YY}),
in the limit of small angles $\theta$, we obtain from
\eqref{rhoGJ} the
following equation for the Goldreich-Julian charge density

\begin{equation}
\label{rhoGJfin} \rho_{\rm{GJ}}=\rho_{\rm{GJ},0}+\delta\rho_{\rm{GJ},\
l^\prime m^\prime}=-\frac{\Omega B_0}{2\pi c }\frac{1}{N\bar{r}
^3}\frac{f(\bar{r} )}{f(1)}\left(1-\frac{\kappa}{\bar{r}
^3}\right)- \frac{1}{4\pi c}\frac{1}{R\bar{r} ^4}\frac{B_0
e^{-i\omega t}}{\Theta^2(\bar{r} )}\frac{1}{N}\frac{f(\bar{r}
)}{f(1)}\tilde{\eta}(\bar{r} )l^\prime(l^\prime+1)Y_{l^\prime
m^\prime} \ ,
\end{equation}
where $\rho_{\rm{GJ},0}$ is the Goldreich-Julian
charge density of a slowly rotating neutron star while
$\delta\rho_{\rm{GJ,\ l^\prime m^\prime}}$ is the
correction induced  by oscillations. We
are here interested in analyzing the Goldreich-Julian charge
density of the first few modes, namely those with $(l^\prime,m^\prime)$ given by
$(0,0)$, $(1,0)$, $(1,1)$, $(2,0)$ and $(2,1)$.
To this extent, however, we
greatly simplify our calculations by
approximating $Y_{l^\prime m^\prime}(\theta,\phi)\approx
A_{l^\prime m^\prime}(\phi)\theta^m$, where the terms $A_{lm}(\phi)$ have
real parts given by
\begin{eqnarray}
\label{listA} \qquad A_{00}=\frac{1}{\sqrt{4\pi}}\
,\qquad A_{10}=\sqrt{\frac{3}{4\pi}}\ ,\qquad
A_{11}=-\sqrt{\frac{3}{8\pi}}\cos\phi\ ,\qquad
A_{20}=\sqrt{\frac{5}{4\pi}}\ ,\qquad
A_{21}=-3\sqrt{\frac{5}{24\pi}}\cos\phi\ .
\end{eqnarray}
From \eqref{rhoGJfin} we can compute the ratio
\be
\label{delta_rhoGJ_2D}
\delta\rho_{\rm{GJ\ l^\prime
    m^\prime}}/\rho_{\rm{GJ},0}=\frac{K}{2\bar{r}^{2-m/2}}\Theta_0^{m-2}\left(\frac{f(\bar{r})}{f(1)}\right)^{\frac{2-m}{2}}\frac{l^\prime
  (l^\prime +1) A_{l^\prime m^\prime}(\phi)}{\left(1-\frac{\kappa}{\bar{r}
^3}\right)} \ ,
\ee
where we have posed
$\tilde{\eta}(\bar{r} )\approx\tilde{\eta}(1)$, which
amounts to the assumption that the
oscillation amplitude maintains the value it has on the
surface of the star, at least within small distances
far from it, as we are considering here. Moreover, we
have introduced the small number
$K=\tilde{\eta}(1)/\Omega R$ to parametrize
the amplitude of the oscillation. Finally, 
have considered $\Theta\approx\theta$.
One can easily see that $\delta\rho_{\rm{GJ\ l^\prime m^\prime}}=0$
for the mode $(0,0)$.
Figure~\ref{rhoGJ_2D_color}, on the other hand,
shows the ratios $\delta\rho_{\rm{GJ\ l^\prime
    m^\prime}}/\rho_{\rm{GJ},0}$ for the other four
modes, computed at $t=0$. When plotting these graphs we have used
the following typical set of parameters: $\kappa=0.15$,
$\varepsilon=1/3$, $K=0.01$, $\Theta_0=0.008$, $\Omega=1 {\rm
rad/s}$\footnote{We
  will show in Sec.\ref{Energy_losses} below that $K$ and
  $\Theta_0$ are not independent of each other and that
  $\Theta_0$ does also depend on the indices $l^\prime$ and
  $m^\prime$. However, we have chosen to use a single
  value $\Theta_0=0.008$ for all of the  plots reported
  in Fig.~\ref{rhoGJ_2D_color} as this nevertheless
  represents a mean value typical of a standard
  astrophysical situation.
}.
As it is clear from Fig.~\ref{rhoGJ_2D_color}, the oscillation induced
Goldreich-Julian charge density not only can be a significant part
of $\rho_{\rm{GJ},0}$, but can even prevail several
hundred times over it,
for example, for the mode $(2,0)$. An exception is
represented by the mode $(1,1)$, for which 
$|\delta\rho_{\rm{GJ\ l^\prime
    m^\prime}}|<|\rho_{\rm{GJ},0}|$ even very close to
the star.
The influence of oscillations
is greater near the surface of the star, which is the most
interesting region of the magnetosphere, while it decreases far
from it.
On the other hand, the relativistic charge density $\rho$ that
enters Eq.~\eqref{Poiss} is proportional to the intensity of the
magnetic field through a proportionality coefficient that is
constant along the given magnetic field line \citep{Muslimov1991},
i.e.
\begin{equation}
\label{rho}
\rho=\frac{\Omega B_0}{2\pi c }\frac{1}{N\bar{r}
^3}\frac{f(\bar{r} )}{f(1)}\left[A(\xi)+e^{-i\omega t}\tilde{a}(
\xi,\phi)\right]\ ,
\end{equation}
where we have introduced the variable $\xi=\theta/\Theta(\bar{r} )$,
and where $A(\xi)$ and $\tilde{a}( \xi,\phi)$ are unknown functions to
be specified from the boundary conditions.
The computation of $A(\xi)$, which
corresponds to the case of pure rotation,
has already been performed
by~\cite{Muslimov1997} (see their Eq.~(58)), showing that
 $A(\xi)\approx\kappa-1$. The computation of $\tilde{a}(
\xi,\phi)$, on the other hand, is
discussed
in Sec.~\ref{Solution_close_to_the_star_surface} below.

%-----------------------------------------------------
\section{Poisson equation}
\label{Poisson_equation}

After inserting
(\ref{rhoGJfin}) and (\ref{rho}) into Eq.~(\ref{Poiss}), we find the
following expression for the Poisson equation
\bea
\label{full_poiss}
&&R^{-2}\left\{N\frac{1}{\bar{r}
^2}\frac{\partial}{\partial\bar{r} }\left(\bar{r} ^2
\frac{\partial}{\partial\bar{r} }\right)+\frac{1}{N\bar{r}
^2\theta}\left[ \frac{\partial}{\partial\theta}\left(\theta
\frac{\partial}{\partial\theta}\right)+\frac{1}{\theta}
\frac{\partial^2}{\partial\phi^2}\right]\right\}\Phi= \nonumber
\\ &&-4\pi\frac{\Omega B_0}{2\pi c
}\frac{1}{N\bar{r} ^3}\frac{f(\bar{r}
)}{f(1)}\left[A(\xi)+e^{-i\omega t}\tilde{a}( \xi,\phi)\right]
-4\pi\frac{\Omega B_0}{2\pi c }\frac{1}{N\bar{r}
^3}\frac{f(\bar{r} )}{f(1)}\left(1-\frac{\kappa}{\bar{r}
^3}\right) - \frac{1}{c}\frac{1}{R\bar{r} ^4}\frac{B_0 e^{-i\omega
t}}{\Theta^2(\bar{r} )}\frac{1}{N}\frac{f(\bar{r}
)}{f(1)}\tilde{\eta}(\bar{r} )l^\prime(l^\prime+1)Y_{l^\prime m^\prime}\ .
\eea
Solving such equation represents a complicated task, and
our strategy consists in, first of all, expanding
the scalar potential $\Phi$ as
\begin{equation}
\label{Phi_sum} \Phi(t,\bar{r} ,\xi,\phi)=\Phi_0(\bar{r}
,\xi)+e^{-i\omega t}\delta\Phi(\bar{r} ,\xi,\phi)\ ,
\end{equation}
where the first term $\Phi_0(\bar{r} ,\xi)$ corresponds to the
case of a non-oscillating but rotating star. In this
case, in fact, we know that  $\Phi_0$ must satisfy
the following equation [see, for example, Eq.~(37) of~\citet{Muslimov1992}]
\begin{equation}
R^{-2}\left\{N\frac{1}{\bar{r} ^2}\frac{\partial}{\partial\bar{r}
}\left(\bar{r} ^2 \frac{\partial}{\partial\bar{r}
}\right)+\frac{1}{N\bar{r} ^2\theta}
\frac{\partial}{\partial\theta}\left(\theta
\frac{\partial}{\partial\theta}\right)\right\}\Phi_0=-\frac{2\Omega
B_0}{c\bar{r} ^3}\frac{f(\bar{r}
)}{f(1)}\left\{\frac{1}{N}A(\xi)+\frac{1}{N}
\left(1-\frac{\kappa}{\bar{r} ^3}\right)\right\} \ .
\end{equation}
The next step consists in expanding in terms of spherical
harmonics  both the second term of (\ref{Phi_sum}), representing
the perturbation to the non-oscillating case, and the function
$\tilde{a}( \xi,\phi)$, which contains the $\phi$ dependence of
the charge density \eqref{rho}. Namely, we write
\bea
\label{series} \delta\Phi(\bar{r}
,\xi,\phi)&=&\sum_{l=0}^{\infty}\sum_{m=-l}^{l}
\delta\Phi_{lm}(\bar{r} )Y_{lm}(\xi,\phi)\ , \\
\label{series_a} \tilde{a}(
\xi,\phi)&=&\sum_{l=0}^{\infty}\sum_{m=-l}^{l}
\tilde{a}_{lm}(\bar{r} )Y_{lm}(\xi,\phi) \ .
\eea
At this point, lengthy but straightforward calculations
allow to derive from Eq.~\eqref{full_poiss} the following equation for
the perturbation $\delta\Phi(\bar{r} ,\xi,\phi)$
\begin{eqnarray}
\label{almost_final_poiss}
R^{-2}N\sum_{l=0}^{\infty}\sum_{m=-l}^{l}\left\{\frac{1}{\bar{r}
^2}\frac{\partial}{\partial\bar{r} }\bar{r} ^2
\frac{\partial}{\partial\bar{r} }-\frac{l(l+1)}{N^2\bar{r}
^2\Theta^2(\bar{r} )}\right\}
\delta\Phi_{lm}Y_{lm}=-4\pi\frac{\Omega B_0}{2\pi c
}\frac{1}{N\bar{r} ^3}\frac{f(\bar{r}
)}{f(1)}\sum_{l=0}^{\infty}\sum_{m=-l}^{l}\tilde{a}_{lm}Y_{lm}
\nonumber
\\ -\frac{1}{c}\frac{1}{R\bar{r} ^4}\frac{B_0}{\Theta^2(\bar{r} )}\frac{1}{N}\frac{f(\bar{r} )}{f(1)}
\tilde{\eta}(\bar{r} )l^\prime(l^\prime+1)Y_{l^\prime m^\prime}\
.
\end{eqnarray}
We now introduce $\delta F_{lm}(\bar{r} )=\bar{r}
\delta\Phi_{lm}(\bar{r} )$, and we exploit the fact that
the spherical harmonics $Y_{lm}$ form a set of orthogonal basis
functions. This means that in
Eq.~\eqref{almost_final_poiss} we must first fix $l^\prime=l$ and
$ m^\prime=m$ and we then equal the coefficients of each
basis function $Y_{lm}$ on the left and on the right hand
side. In this way we finally obtain the
Poisson equation in the form
\begin{equation}
\label{final_poiss}
\frac{R^{-2}N}{\bar{r} }\left\{\frac{d^2}{d\bar{r}
^2}-\frac{l(l+1)}{N^2\bar{r} ^2\Theta^2(\bar{r} )}\right\} \delta
F_{lm}(\bar{r} )=-\frac{ B_0}{cN\bar{r} ^3}\frac{f(\bar{r}
)}{f(1)}\left\{2\Omega \tilde{a}_{lm}+ \frac{1}{R\bar{r}
}\frac{\tilde{\eta}(\bar{r} )l(l+1)}{\Theta^2(\bar{r} )}\right\}\
.
\end{equation}
It is worth stressing that Eq.~\eqref{final_poiss} is
valid for small polar angles $\theta$, but for any
distance $\bar{r}$ from the star surface. The limit of
Eq.~\eqref{final_poiss} for small $\bar{r}$ is considered
in the rest of this Section.

%------------------------------------------------------
\subsection{Solution close to the star surface}
\label{Solution_close_to_the_star_surface}

As a first example, we wish to compute the solution of the
Poisson equation close to the star surface,
where $z=\bar{r} -1\ll1$, while imposing no restrictions
on the aperture angle $\Theta_0$
of the last closed magnetic field line.
In this case Eq.~\eqref{final_poiss} for the unknown
$\delta F_{lm}$ becomes
\begin{equation}
\label{final_poiss_polar}
\left[\frac{d^2}{dz^2}-\frac{l(l+1)}{(1-\varepsilon)\Theta^2_0}\right]
\delta F_{lm}(z)=-\frac{2\Omega
B_0R^2}{c}\frac{1-2z}{1-\varepsilon}\tilde{a}_{lm}- \frac{B_0
R}{c\Theta^2_0}\frac{1-3z}{1-\varepsilon}\tilde{\eta}(1)l(l+1)\ ,
\end{equation}
which represents the extension of Eq. (44)
of~\citet{Muslimov1992} to account for the presence of oscillations.
The general solution of Eq.~\eqref{final_poiss_polar}, which
is an  inhomogeneous differential equation, is
\begin{equation}
\delta
F_{lm}=\tilde{C}e^{-\frac{\sqrt{l(l+1)}}{\sqrt{1-\varepsilon}\Theta_0}z}
+\frac{\Theta_0^2}{l(l+1)}\frac{B_0R}{c}\left\{2\Omega
R(1-2z)\tilde{a}_{lm}+\frac{l(l+1)}{\Theta_0^2}\tilde{\eta}(1)(1-3z)\right\}
\ .
\end{equation}
The constant $\tilde{C}$ and the coefficients
$\tilde{a}_{lm}$ may be found after
imposing physically motivated boundary conditions.
In particular, at the star surface we
require both equipotentiality and absence of a steady
state electric field, which amounts to the two conditions
\begin{equation}
\delta F_{lm}|_{z=0}=0\ ,\ \frac{d\delta F_{lm}}{d
z}\bigg\vert_{z=0}=0 \ .
\end{equation}
From them, simple calculations allow to derive
\begin{equation}
\label{almtildec} \tilde{a}_{lm}=-\frac{l(l+1)}{2\Omega
R\Theta_0^2}\tilde{\eta}(1)\left[1-\frac{
\Theta_0\sqrt{1-\varepsilon}}{\sqrt{l(l+1)}
-2\Theta_0\sqrt{1-\varepsilon}}\right]
,\qquad\qquad\qquad
\tilde{C}=-\frac{B_0R}{c}\tilde{\eta}(1)\frac{\Theta_0\sqrt{1-\epsilon}}
{\sqrt{l(l+1)}-2\Theta_0\sqrt{1-\epsilon}} \ .
\end{equation}
The final solution for the scalar potential near the surface
of the oscillating rotating neutron star has
therefore the form
\be
\label{close_to_surface0}
\Phi(t,\bar{r} ,\xi,\phi)=\Phi_0(\bar{r} ,\xi)+e^{-i\omega
t}\frac{1}{\bar{r}
}\frac{B_0R}{c}\tilde{\eta}(1)\sum_{l=0}^{\infty}\sum_{m=-l}^{l}
\frac{\Theta_0\sqrt{1-\epsilon}}
{\sqrt{l(l+1)}-2\Theta_0\sqrt{1-\epsilon}}
\Bigg\{
-e^{-\frac{\sqrt{l(l+1)}}{\sqrt{1-\varepsilon}\Theta_0}(\bar{r}
-1)}+1 -\frac{\sqrt{l(l+1)}}{\sqrt{1-\varepsilon}\Theta_0}
(\bar{r} -1)\Bigg\}Y_{lm}(\xi,\phi) \ee
and the corresponding expression for the accelerating component of
the electric field is
\begin{equation}
\label{close_to_surface1}
E_{\|}=E_0-e^{-i\omega t}\frac{B_0R}{c}\tilde{\eta}(1)
\sum_{l=0}^{\infty}\sum_{m=-l}^{l}\frac{\sqrt{l(l+1)}}{\sqrt{l(l+1)}-2\sqrt{1-\varepsilon}\Theta_0}
\left\{e^{-\frac{\sqrt{l(l+1)}}{\sqrt{1-\varepsilon}\Theta_0}(\bar{r}
-1)}-1\right\} Y_{lm}(\xi,\phi)\ ,
\end{equation}
where
\begin{equation}
E_0=-\frac{1}{R}\frac{\partial\Phi_0}{\partial\bar{r} }\ ,
\end{equation}
is the accelerating field that is present even
in the absence of  oscillations.

%---------------------------------------------------
\subsection{Solution in the polar cap region}
\begin{figure*}
\begin{center}
\includegraphics[width=8.5cm,angle=0]{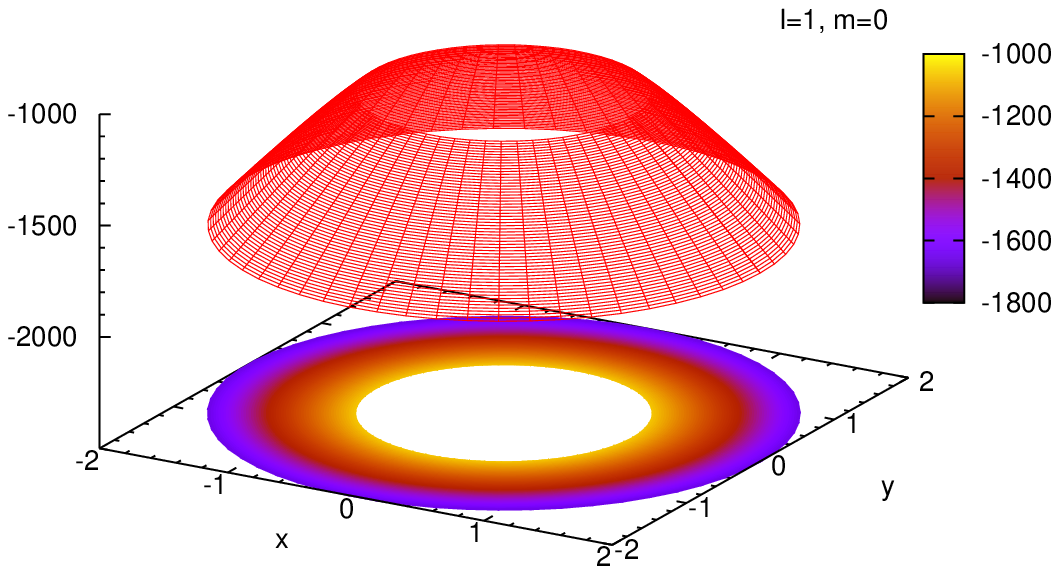}
\hspace{0.125truecm}
\includegraphics[width=8.5cm,angle=0]{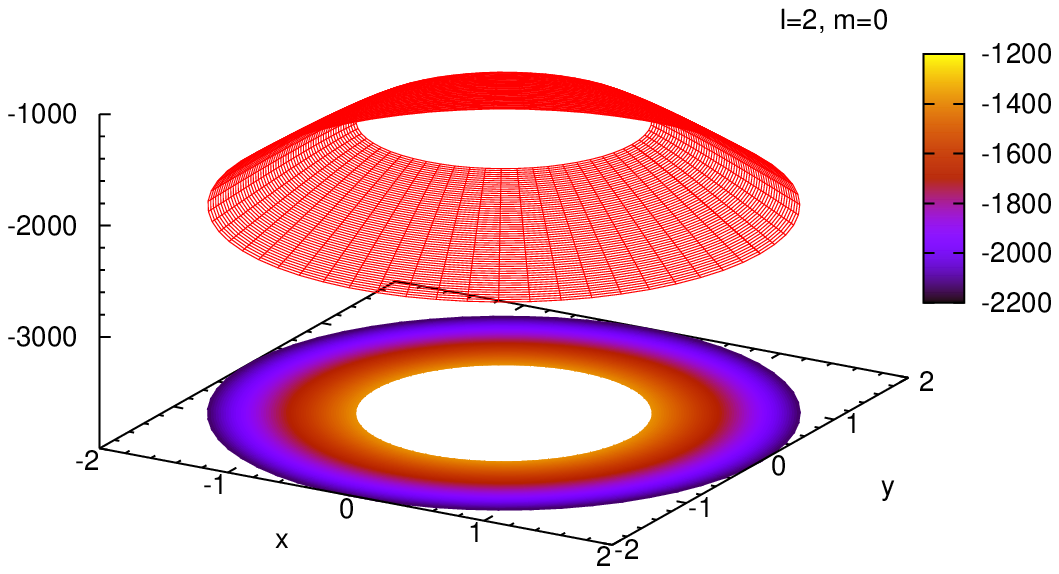}
\includegraphics[width=8.5cm,angle=0]{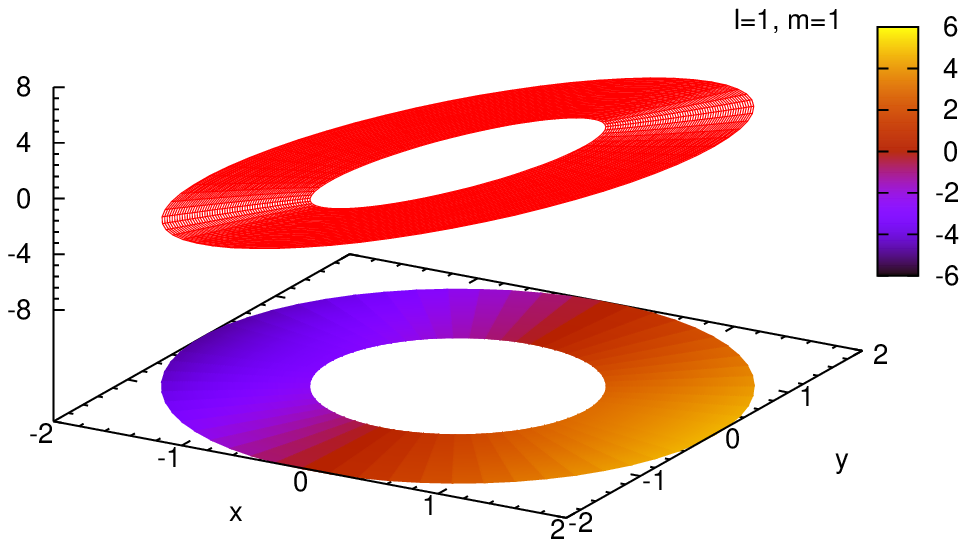}
\hspace{0.125truecm}
\includegraphics[width=8.5cm,angle=0]{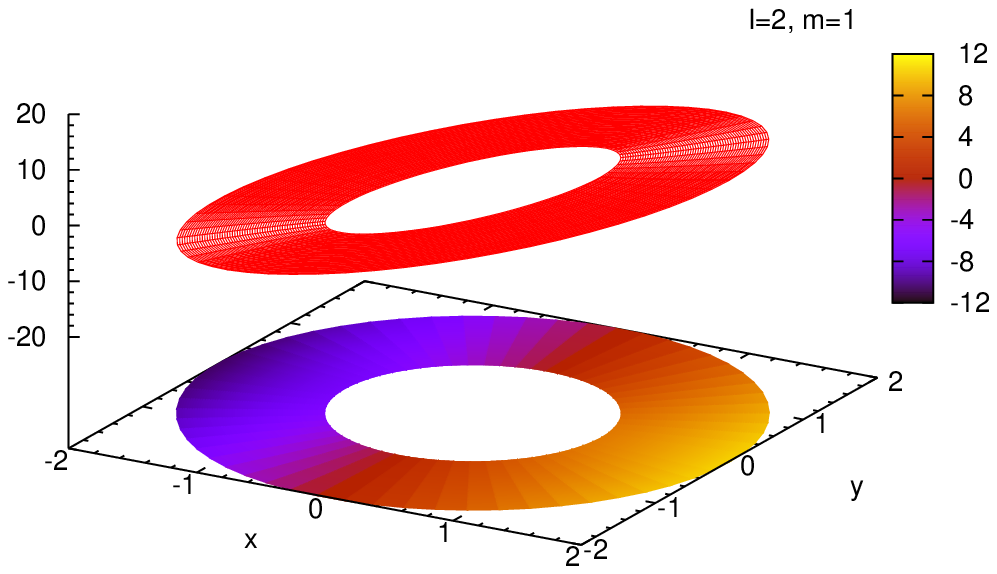}
\end{center}
\vspace*{-0.7cm}
\caption{Ratio ${\delta E_{\ lm}}/{E_{0}}$
for the mode $(1,0)$
(left top panel),  for the mode $(1,1)$ (left bottom
  panel), for the mode  $(2,0)$ (right top panel) and for
  the mode $(2,1)$ (right bottom panel).
}
\label{figE}
\end{figure*}

The solution in the polar cap region, namely
when  $\Theta_0\ll\bar{r}-1\ll R_{LC}/R$, can be obtained
from Eq.~\eqref{close_to_surface0} and
Eq.~\eqref{close_to_surface1} in the limit of small
$\Theta_0$, or, directly, from
Eq.~\eqref{final_poiss}. In this case, in fact,
$|d^2\delta F_{lm}/d\bar{r} ^2|\ll l(l+1)|\delta
  F_{lm}|/N^2\bar{r}^2\Theta^2(\bar{r})$, and
Eq.~\eqref{final_poiss} therefore reduces to
\begin{equation}
\frac{l(l+1)}{R^2N\bar{r} ^3\Theta^2(\bar{r} )}\delta
F_{lm}=\frac{2\Omega B_0}{c}\frac{1}{N\bar{r} ^3}\frac{f(\bar{r}
)}{f(1)}\tilde{a}_{lm}+\frac{1}{c}\frac{B_0}{\Theta^2(\bar{r} )}
\frac{1}{RN\bar{r} ^4}\frac{f(\bar{r} )}{f(1)}\tilde{\eta}(\bar{r}
)l(l+1) \ .
\end{equation}
After using $\Theta$ as given by Eq.~(\ref{Theta})
and with the same coefficients $\tilde{a}_{lm}$ expressed
by Eq.~(\ref{almtildec}), we  can obtain

\begin{equation}
\delta F_{lm}=\frac{B_0R}{c}\left[-\bar{r}
\tilde{\eta}(1)\frac{\sqrt{l(l+1)}-3\sqrt{1-\varepsilon}\Theta_0}
{\sqrt{l(l+1)}-2\sqrt{1-\varepsilon}\Theta_0}+\frac{f(\bar{r}
)}{f(1)} \frac{\tilde{\eta}(\bar{r} )}{\bar{r} }\right]\ .
\end{equation}
This allows to derive both the electric
potential and the accelerating component of
the electric field as\footnote{
For small values of $\Theta_0$ we use the approximation
$(\sqrt{l(l+1)}-3\sqrt{1-\varepsilon}\Theta_0)/(\sqrt{l(l+1)}-2\sqrt{1-\varepsilon}\Theta_0)\sim1$.}
\bea
\label{polar_cap0}
\Phi(t,\bar{r} ,\xi,\phi)&=&\Phi_0(\bar{r} ,\xi)+
\frac{B_0R}{c}\left[-\tilde{\eta}(1)+\frac{f(\bar{r}
)}{f(1)} \frac{\tilde{\eta}(\bar{r} )}{\bar{r}
^2}\right]\sum_{l=0}^{\infty}\sum_{m=-l}^{l}Y_{lm}(\xi,\phi)\ , \\
\label{polar_cap1}
E_{\|}&=&E_0-\frac{B_0}{c}\frac{d}{d\bar{r}
}\left(\frac{f(\bar{r} )}{f(1)} \frac{\tilde{\eta}(\bar{r}
)}{\bar{r}
^2}\right)\sum_{l=0}^{\infty}\sum_{m=-l}^{l}Y_{lm}(\xi,\phi)\ .
\eea
It is interesting to compare the second term on the right
hand side of Eq.~\eqref{polar_cap1}, namely the contribution
due to the oscillations, with
$E_0$ as reported, for example, in Eq.~(55)
of \citet{Muslimov1992}. We find
\begin{equation}
\label{ratio_of_E} \frac{\delta E_{\ lm}}{E_{0}}=e^{-i\omega
t}\frac{2}{3}\frac{\tilde{\eta}(1)}{\Omega R\kappa}
\left[\frac{d}{d\bar{r}}\left(\frac{f(\bar{r})}{f(1)}\frac{1}{\bar{r}^2}
\right)\right]\bar{r}^4\Theta_0^{m-2}\frac{\xi^m}{1-\xi^2}A_{lm}(\phi)\ ,
\end{equation}
that we plot in Fig.~\ref{figE}  for the different modes considered so
far and computed at $t=0$ using the same set of parameters as for
the plots in Fig~\ref{rhoGJ_2D_color}. Moreover, we have posed $\xi=1/2$,
i.e.  we have considered the middle magnetic field line between
the polar axis and the edge of the polar cap. The plots
demonstrate the significant influence of oscillations to the
electric field of a pulsar. In particular, because $\Theta_0$ is
supposed to be small, the ratio $\delta E_{\ lm}/E_{0}$ can be
very large for modes with $m<2$. Indeed, the modes
$(l,m)=(0,0),(2,0)$  induce an electric field in the opposite
direction to $E_0$, which is three orders of magnitude larger in
modulus than  $E_0$. In general, the ratio $|\delta E_{\ lm}/E_{0}|$
increases when increasing the distance from the star surface.

In a recent work aimed at quantifying the impact of neutron star
oscillations on the accelerating electric field,
\citet{Timokhin2007} considered the whole range of angles $\theta$
and more complicated cases of spheroidal oscillations. He found
that the contribution of oscillations to such electric field may
be either positive or negative and that this contribution is
substantial for values of $l$ of several hundreds. Indeed, as
shown in the Sec.~\ref{Energy_losses} below, the appearance of the
small parameter $\Theta_0$ in the computation of the energy losses
makes astrophysically relevant even the modes with small values of
$l$ and $m$.

%
%\begin{figure}
%\includegraphics[width=7.cm,angle=0]{....ps}
%\caption{The ratio $L_{m\neq0}/L_{rot}$ as a function of $\eta$
%for modes $(1,1)$ (red line) and $(2,1)$ (blue line).}
%\label{figE}
%\end{figure}
%

%====================================================
\section{Energy losses}
\label{Energy_losses}

In this Section we calculate the energy losses from the polar cup
region of a rotating and oscillating neutron star, using several
results presented in~\cite{Abdikamalov2009} for the spherical
Schwarzschild star. The total energy loss from the open field
lines region, averaged over an oscillation period and carried out
by the outflowing plasma, is determined as \citep{Timokhin2000}
\begin{equation}
\label{totalL}
L_{lm}=\frac{1}{\tau}\int_0^{\tau}dt \int_0^{2\pi}d\phi\int_0^{\Theta_0}d\theta\vert
j^{\hat{r}}_{lm}(R,\theta,\phi)
\Delta\varepsilon_{lm}(\theta,\phi)\vert R^2\sin\theta\ ,
\end{equation}
where $\Delta\varepsilon(\theta,\phi)$ is the work done by the
electric field to move a unit charge to the point with coordinate
$(R,\theta,\phi)$
\begin{equation}
\label{deltavarepsilon}
\Delta\varepsilon_{lm}=RN_R^2\int_0^{\theta}E^{\hat{\theta}}_{GJ\
lm}(R,\theta',\phi) d\theta' \ ,
\end{equation}
while $j^{\hat{r}}_{lm}$ is the electric current density and
$\tau$ is the period of oscillations. It should be noted that
despite some terms in the expression for the energy losses will
contain the factor $e^{-i\omega
  t}$, which mathematically gives a
zero net result when averaged over time, physically these
terms will give a non zero net contribution.
That is because the average crossing time of the acceleration zone by
particles which are responsible for the energy losses is
much shorter than the oscillation period (typical frequencies of
oscillations are of the order of $10-50 kHz$).
In practice, therefore, the particles
do not return back to the star (see also
the discussion of \cite{Timokhin2000}). If the polar cap region is in the
conditions of complete charge separation, the current density is
well approximated as \citep{Ruderman1975, Timokhin2000}
\begin{equation}
\label{j} j^{\hat{r}}\simeq\rho_{GJ}c\ .
\end{equation}
In order to compute the integral in \eqref{totalL}, we
need to provide an estimate of the
angle $\Theta_0$ of the last closed magnetic field line at the surface of
the star. After using
$R_a$ to denote the radial coordinate
of the point where the last closed magnetic field line crosses the
equatorial plane (note that $R_a\gg R$),
we derive $\Theta_0$ from the condition
\begin{equation}
\label{alfven}
\epsilon_{pl}(R_a,\pi/2,\phi)=\epsilon_{em}(R_a,\pi/2,\phi)
\ ,
\end{equation}
where $\epsilon_{pl}$ and $\epsilon_{em}$ are the
kinetic energy density of the outflowing plasma and the energy
density of the magnetic field, respectively.
The underlining idea, in fact, is that at the last closed
magnetic field line equipartition of energy exists between
the magnetic field and the plasma, and, for convenience, the condition is
evaluated at the equator.
The two energies $\epsilon_{pl}$ and $\epsilon_{em}$ have
already been computed by~\citet{Abdikamalov2009} [see
their Eq. (100) and (101)] and are
\bea
\label{epl}
\epsilon_{pl}(R_a,\pi/2,\phi)&=&\frac{1}{2f(1)}\frac{N_R}{N_{R_{a}}}\frac{R^3}{R_a^3}
\Delta\varepsilon \ j^{\hat{r}}\ , \\
\label{eem}
\epsilon_{em}(R_a,\pi/2,\phi)&=&\frac{N_{R_{a}}}{32\pi}\frac{R^6}{R_a^6}B_0^2\
,
\eea
where $N_R=\sqrt{1-2M/R}$,
\ $N_{R_{a}}=\sqrt{1-2M/R_a}\approx1$. The angle
$\Theta_0$ does not explicitly appear in \eqref{epl} and
\eqref{eem}, but rather implicitly. In fact,
for a dipole magnetic field $f(r)\sin^2\theta/r=\rm{const}$ and therefore
\begin{equation}
\frac{R}{R_a}=\frac{f(1)}{f(R_a/R)}\Theta_0^2\ ,
\end{equation}
where $f(R_a/R)$ is close to unity. In order to proceed
with the computation of $\Theta_0$, and hence of the energy loss,
we need the electric field component
\begin{eqnarray}
\label{EGJtheta} (E^{\hat{\theta}})_{GJ\ lm} &=&
-\frac{1}{Nc}\left[\left(1-\frac{\kappa}{\bar{r} ^3}\right)\vec{u}
+\vec{\delta v}\right]_{\hat{\phi}}B_{\hat{r}}\nonumber\\
&=&-\frac{1}{Nc}\left[\Omega R\bar{r}
\sin\theta\left(1-\frac{\kappa}{\bar{r}
^3}\right)-\partial_{\theta}Y_{lm}
(\theta,\phi)\tilde{\eta}(\bar{r} )\right]B_0\frac{f(\bar{r}
)}{f(1)}\frac{1}{\bar{r} ^3}\cos\theta \ ,
\end{eqnarray}
from which we can compute $\Delta\varepsilon_{\ lm}$ through
\eqref{deltavarepsilon}.
This allows us to obtain
(for small angles $\theta$)
\bea
\Delta\varepsilon_{\ lm}&=&-\frac{RN_RB_0}{c}\left[\Omega
R(1-\kappa)\frac{\theta^2}{2}-A_{lm}\theta^m\tilde{\eta}(1)\right]
\ \ \ \ \ \ \ \ \ \rm{for} \ \ \ \ m\neq0 \ , \\
%in the case $m\neq0$ and
\Delta\varepsilon_{\ lm}&=&-\frac{RN_RB_0}{c}\frac{\theta^2}{2}\left[\Omega
R(1-\kappa)-2 A_{l0}\tilde{\eta}(1)\right] \ \ \
\ \ \ \ \ \ \ \ \ \  \ \ \ \rm{for}  \ \ \ \ m=0 \ .
\eea
Taking into account equations (\ref{alfven}),
(\ref{epl}) and (\ref{eem}), we can obtain the expression
for the angle $\Theta_0$, which is  an algebraic equation
in the case $m\neq0$ while it has a closed form in the
case $m=0$, i.e.
\bea &&\label{theta_0_mneq0}
\frac{RN_R}{c}\left[\Omega
R(1-\kappa)\frac{\Theta_0^2}{2}-A_{lm}\Theta_0^m\tilde{\eta}(1)\right]\left\{
\frac{\Omega
(1-\kappa)}{c}+\frac{\tilde{\eta}(1)}{2c R}l(l+1)A_{lm}\Theta_0^m\right\}=
\frac{1}{8}f^4(1)\Theta_0^6 \hspace{1.1cm} \ \ \rm{for} \ \ \ \ m\neq0 \ ,  \\
\label{theta0}
&&\Theta_{0}=\frac{2}{f(1)}\frac{\sqrt[4]{R}\sqrt[4]{N_R}}{\sqrt[4]{2}\sqrt[4]{c}}\left[\Omega
R(1-\kappa)-2A_{l0}\tilde{\eta}(1)\right]^{1/4}\left\{
\frac{\Omega}{2c}(1-\kappa)+\frac{1}{4c}\frac{1}{R}\tilde{\eta}(1)l(l+1)A_{l0}\right\}^{1/4}
\ \ \ \ \ \ \ \ \ \ \  \ \ \ \rm{for}  \ \ \ \ m=0 \ . \eea
When $m=0$ and $\Omega=0$, hence in the axisymmetric case with no
rotation, we get the expression
\begin{equation}
\Theta_0\vert_{m=0,\Omega=0}=\sqrt[4]{\frac{4A_{l0}^2l(l+1)N_R\tilde{\eta}^2(1)}{f^4(1)}}\
.
\end{equation}
This estimate is slightly different from that obtained by~\citet{Abdikamalov2009}
\begin{equation}
\Theta_0=2N_R^{1/4}\left[\frac{\tilde{\eta}(1)A_{l0}^{(2)}}{f(1)}\right]^{1/2}\
,
\end{equation}
firstly because of the different expansion of the spherical
harmonics, which in~\citet{Abdikamalov2009} was chosen to be
$Y_{lm}(\theta,\phi)\approx
A_{lm}^{(1)}(\phi)\theta^m+A_{lm}^{(2)}(\phi)\theta^{m+2}$, and,
secondly, because of the different determination of $B_0$
(equations (73)-(75) for magnetic field in the paper
of~\citet{Abdikamalov2009} do not contain $f(1)$ in the
denominator). On the other hand, when $m=0$ and $\tilde\eta=0$,
hence in the axisymmetric case with no oscillations, we get the
expression,
\begin{equation}
\label{theta_rot} \Theta_0\vert_{m=0,
\tilde\eta=0}=\sqrt[4]{\frac{4N_R}{f^4(1)}\frac{\Omega^2R^2}{c^2}(1-\kappa)^2}\
,
\end{equation}
which represents the correct general relativistic
extension of the expression reported by~\citet{Muslimov1992}.
Inserting the angles $\Theta_0$ defined by the equations
(\ref{theta_0_mneq0}) and (\ref{theta0}) into
(\ref{totalL}), we
derive the total energy losses from the polar cap region for
a rotating and oscillating magnetized neutron star as
\begin{eqnarray}
\label{Lmneq0}
L\vert_{m\neq0}&=&\int_0^{2\pi}d\phi\frac{R^3N_RB_0^2}{2\pi}\Bigg|\Bigg\{\frac{\Omega^2R}{2
cN_R}(1-\kappa)^2\frac{\Theta_0^3}{3}+\frac{\Omega}{4
c}\frac{1}{N_R}(1-\kappa)\tilde{\eta}(1)l(l+1)A_{lm}(\phi)\frac{\Theta_0^{m+4}}{m+4}
\nonumber \\ &&-\frac{\Omega}{2
c}\frac{1}{N_R}(1-\kappa)A_{lm}(\phi)\tilde{\eta}(1)\frac{\Theta_0^{m+2}}{m+2}-\frac{1}{2
c}\frac{1}{RN_R}A^2_{lm}(\phi)\tilde{\eta}^2(1)l(l+1)\frac{\Theta_0^{2m+2}}{2m+2}\Bigg\}\Bigg|
\hspace{2.8cm}{\rm
  for} \ \ \ \ \ m\neq0 \ , \\
\label{Lm=0}
L\vert_{m=0}&=&\int_0^{2\pi}d\phi\frac{R^3N_RB_0^2}{2\pi}\frac{\Theta_0^2}{8}\Bigg|\left[\Omega
R(1-\kappa)-2A_{l0}(\phi)\tilde{\eta}(1)\right]\left\{
\frac{\Omega}{cN_R}(1-\kappa)+\frac{1}{2c}\frac{1}{N_R}\tilde{\eta}(1)l(l+1)A_{l0}(\phi)\right\}\Bigg|
\hspace{0.4cm}{\rm
  for} \ \ \ \ \ m=0 \ .
\eea
%
%In Fig.~\ref{fig2} one can find the radial dependence of parallel
%electric field $E_{\|}$ in terms of $E_{vac}$ for the different
%values of ... parameter.
Equations \eqref{Lmneq0} and \eqref{Lm=0} just derived
are astrophysically very relevant and deserve some
comments. In the first place, it is interesting to note
that (\ref{Lm=0}) takes a simpler form if only the linear
terms in  the amplitude of the stellar oscillation are
retained. In this case, in fact, we find
\begin{equation}
\label{Lm=0b}
L\vert_{m=0}=\frac{R^4B_0^2\Omega^2}{8c}(1-\kappa)^2\Theta_0^4\left[1+
\frac{\tilde{\eta}(1)}{\Omega
R}\frac{A_{l0}}{1-\kappa}\frac{l^2+l-2}{2}\right]\ .
\end{equation}
We can highlight the corrections  with respect to the
energy losses in the Newtonian case and in the absence of
oscillations if we replace $\Theta_0$ in \eqref{Lm=0b}
with one of the expressions computed above. For simplicity, we
consider the case given by (\ref{theta_rot}), namely the
case of pure rotation, and we obtain
\begin{equation}
\label{Lm=01} L\vert_{m=0}=3(1-\kappa)^4\frac{N_R}{f^4(1)}\left[1+
\frac{\tilde{\eta}(1)}{\Omega
R}\frac{A_{l0}}{1-\kappa}\frac{l^2+l-2}{2}\right](\dot{E}_{rot})_{Newt}\
,
\end{equation}
where $(\dot{E}_{rot})_{{\rm Newt}}$ is the standard Newtonian expression
for the magneto-dipole losses in flat space-time approximation
\begin{equation}
(\dot{E}_{rot})_{{\rm Newt}}=\frac{1}{6}\frac{\Omega^4B_0^2R^6}{c^3}\ .
\end{equation}
The astrophysical relevance of Eq.~(\ref{Lm=01}) becomes even more
transparent when it is rewritten in terms of the pulsar
observables $P$, i.e. the period, and $\dot{P}\equiv dP/dt$. To
this extent we first recall two standard relations of pulsar
physics. The first one is the relation between luminosity and
spin-down, namely
\begin{equation}
\label{LPP} L=-\tilde{I}\Omega\dot{\Omega}
\end{equation}
where $\tilde{I}$ is the general relativistic
moment of inertia of the star ~\citep{Rezzolla2004}.
The second (Newtonian)
relation is between the spin down and the intensity
of the magnetic field, i.e.
\begin{equation}
(P\dot{P})_{Newt}\equiv\left(\frac{2\pi^2}{3c^3}\right)
\frac{R^6B^2_0}{I} \ ,
\end{equation}
where $I$ is the classical moment of inertia of the star.
From these two relations  and from Eq.~(\ref{Lm=01}) we
deduce
\begin{equation}
\label{PP} (P\dot{P})_{max}=\frac{3}{2}(1-\kappa)^4\left[1+
\frac{\tilde{\eta}(1)}{\Omega
R}\frac{A_{l0}}{1-\kappa}\frac{l^2+l-2}{2}\right]\frac{I}{\tilde{I}}
\frac{N_R}{f^4(1)}(P\dot{P})_{Newt}\ .
\end{equation}
Pulsar periods $P$ and spin-down rates $\dot P$
are very precisely measured quantities for a large number
of pulsars.
\citet{Kaspi2004}, for instance, report a $P-\dot{P}$ diagram of
1403 cataloged rotation-powered pulsars (see also~\cite{Arons2007}).
Thus, expression (\ref{PP}) for $P\dot{P}$ can in
principle be applied to existing observations
for detecting the possible effects of stellar surface oscillations.
The main difficulty encountered in this kind of
analysis rests in the low accuracy measurements of the
stellar radius, which reflects in poor estimates
of the moment of inertia.
However, the predictive potentiality of \eqref{PP} is
unquestionable, and once the moment of inertia of
neuron stars is determined more precisely, it will
provide detailed information about stellar oscillations
starting  from fundamental observational properties about pulsar timing.

We now discuss the energy losses in the case $m\neq0$, and to this
extent we rewrite Eq.~(\ref{Lmneq0}) as
\begin{eqnarray}
\label{Lmneq0_for_graph}
L_{m\neq0}&=&L_{rot}\int_0^{2\pi}d\phi\frac{1}{2\pi}\Bigg|\left\{1+K\frac{2}{1-\kappa}l(l+1)A_{lm}(\phi)\frac{\Theta_0^m}{m+4}
-K\frac{8}{1-\kappa}A_{lm}(\phi)\frac{\Theta_0^{m-2}}{m+2}\right\}\Bigg|
\nonumber \\ &&
\qquad\qquad\qquad-\frac{R^4B_0^2\Omega^2}{8c}(1-\kappa)^2\Theta_0^4
\int_0^{2\pi}d\phi\frac{1}{2\pi}\Bigg|K^2
4l(l+1)A_{lm}^2(\phi)\frac{\Theta_0^{2m-2}}{2m+2}\Bigg|\ ,
\end{eqnarray}
where $L_{rot}$ denotes pure rotational energy losses
\begin{equation}
L_{rot}=\frac{R^4B_0^2\Omega^2}{8c}(1-\kappa)^2\Theta_0^4
\ .
\end{equation}
We are here interested
again in analyzing the energy losses of the
first few modes, namely those with $(l,m)$ given by $(0,0)$,
$(1,1)$, $(2,0)$ and $(2,1)$.
Eq.~(\ref{Lmneq0_for_graph}) contains
two major contributions. The first contribution includes the
rotational energy losses plus
terms linear in $K$ which are not present neither in pure
rotational nor in pure oscillatory regime. The second
contribution, on the other hand,
is the only one present in the pure oscillatory regime.
More specifically, in
the pure rotation regime ($\tilde{\eta}=0$, $\Omega\neq 0$), only
the first term in the curl brackets of
Eq.~\eqref{Lmneq0_for_graph} survives, and we simply
recover $L=L_{rot}$.
In the opposite regime of
pure oscillation ($\tilde{\eta}\neq 0$, $\Omega = 0$) the
whole first contribution to the right hand side of
Eq.~\eqref{Lmneq0_for_graph} vanishes and only the second
contribution proportional to $K^2$ survives. Finally, in the mixed regime with
both $\tilde{\eta}\neq 0$ and $\Omega\neq 0$, all of the
terms must be included.

In order to appreciate the dependence of the energy losses on the
oscillation amplitudes, we have computed the ratio
$L_{m\neq0}/L_{rot}$ as a function of $K$ for all of the modes
mentioned above. As typical and representative parameters of the
rotating neutron star we have chosen $R=10 {\rm km}$, $\Omega=1
{\rm rad/s}$, $\varepsilon=1/3$. We first computed the angle
$\Theta_0$ after applying a standard root solver to
Eq.~(\ref{theta_0_mneq0}). Graphs of $\Theta_0$ as a function of
the parameter $K$ are presented in Fig.~\ref{Theta0} for the modes
$(1,1)$ and $(2,1)$ and they show that the size of polar cap
increases with the amplitude of stellar oscillations.
Fig.~\ref{fig4}, on the other hand, reports the ratio
$L_{m\neq0}/L_{rot}$. Interestingly, it follows from
Eq.~(\ref{Lmneq0_for_graph}) that modes with $m=1$ have the small
parameter $\Theta_0$ to a negative power. Therefore, for such
modes the energy losses due to oscillations may exceed
significantly the energy losses due to pure rotation, even for
relatively small $K$. This effect is indeed shown in the left
panel of Fig.~\ref{fig4} which reports the ratio
$L_{m\neq0}/L_{rot}$ for the modes $(1,1)$ and $(2,1)$. The right
panel of Fig.~\ref{fig4}, on the other hand, reports the ratio
$L_{m\neq0}/L_{rot}$ for the modes $(0,0)$ and $(2,0)$\footnote{
The mode $(1,0)$ does not have linear contributions in $K$ to the
energy losses, as it is follows from (\ref{Lm=01}).} and they turn
out to be a factor ten smaller than the ones reported in the left
panel, for the reason explained above. We note, on the contrary,
that the energy losses due to oscillations are smaller than those
due to rotation for the mode $(0,0)$. Modes with $m>1$ do not
contain $\Theta_0$ to a negative power. Although modes with higher
$m$ practically do not give any contribution to the energy losses,
we should remark that for modes with $m>3$ the angle $\Theta_0$
may not be small, thus requiring an alternative approach to the
one presented in this paper.
\begin{figure}
\begin{center}
\includegraphics[width=7.cm,angle=0]{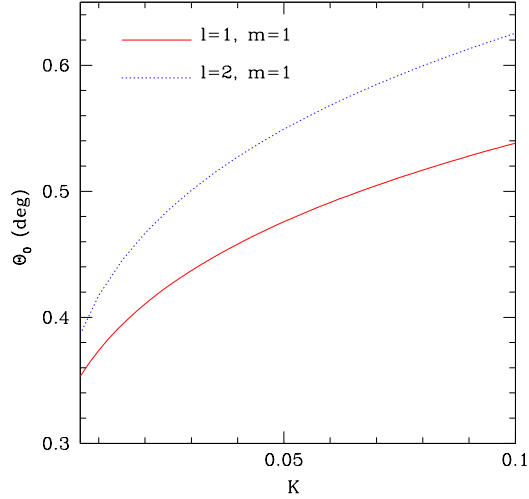}
\caption{The angle (in deg units) of the last open
  magnetic field line $\Theta_0$ as a function of the
  parameter $K=\tilde{\eta}(1)/\Omega R$ for modes
  $(1,1)$ (continuous red line) and $(2,1)$ (dotted blue line).
The representative parameters of the star are $R=10 {\rm km}$,
$\Omega=1 {\rm rad/s}$, $\varepsilon=1/3$. } \label{Theta0}
\end{center}
\end{figure}
Since the energy losses of the pulsar are proportional
to the spin down rate, as it is clear from Eq.~\eqref{LPP},
the same graphs obtained for the
ratios $L_{m}/L_{rot}$  describe also the ratio
$\dot{\Omega}_{lm}/\dot{\Omega}_{rot}$, where $\dot{\Omega}_{lm}$
denotes the time derivative of the pulsar rotation frequency when
the star oscillates with the mode $(l,m)$ while $\dot{\Omega}_{rot}$
corresponds to the case of pure rotation.
Thus, it follows from
Fig.~\ref{fig4} that an observable value of
$\dot{\Omega}_{lm}/\dot{\Omega}_{rot}\sim1.5$ may be reached for
a value of $K$ as small as $K\sim0.03$ for the mode $(1,1)$ and for
$K\sim0.01$ for the mode $(2,1)$.

Finally, it is interesting to
note that the comparison of our results with those of
\cite{Timokhin2000} can be done only with caution, since he
considered the whole range of angles $\theta$, more complicated
cases of spheroidal oscillations, and he did not show the ratio
$L_{m\neq0}/L_{rot}$, but rather the ratio between plasma energy
losses and vacuum energy losses, both in the presence of
oscillations\footnote{What he found is that for some
  modes, like for instance $(1,0)$ or $(l\geq 2,m=0,1)$
  the plasma energy losses are larger than the vacuum
  ones even for small oscillation amplitudes.}.
It should also be noted that \cite{Timokhin2000} performed 
estimations rather than exact calculations.

\begin{figure}
\begin{center}
\includegraphics[width=7.cm,angle=0]{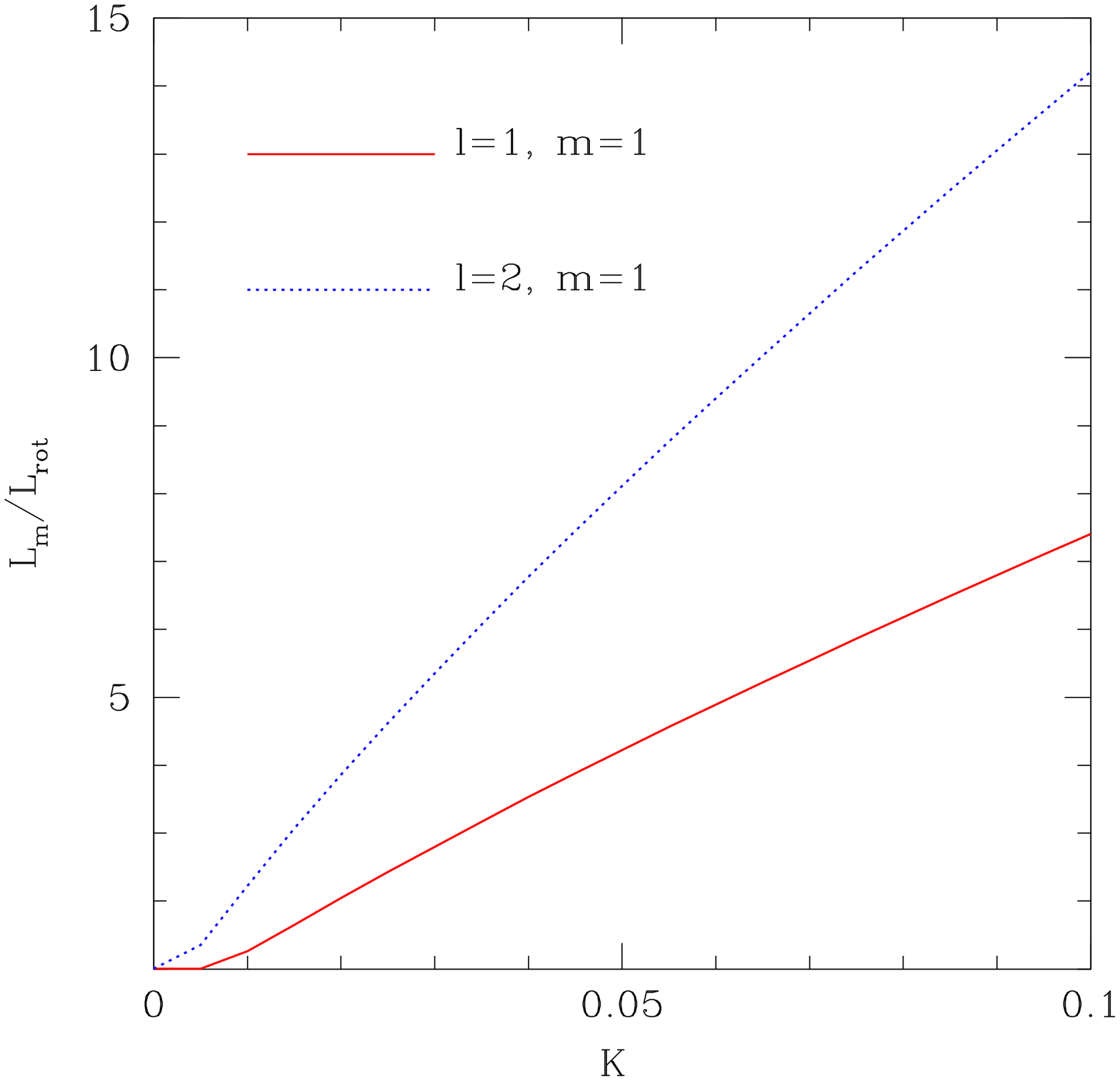}
\hspace{0.125truecm}
\includegraphics[width=7.cm,angle=0]{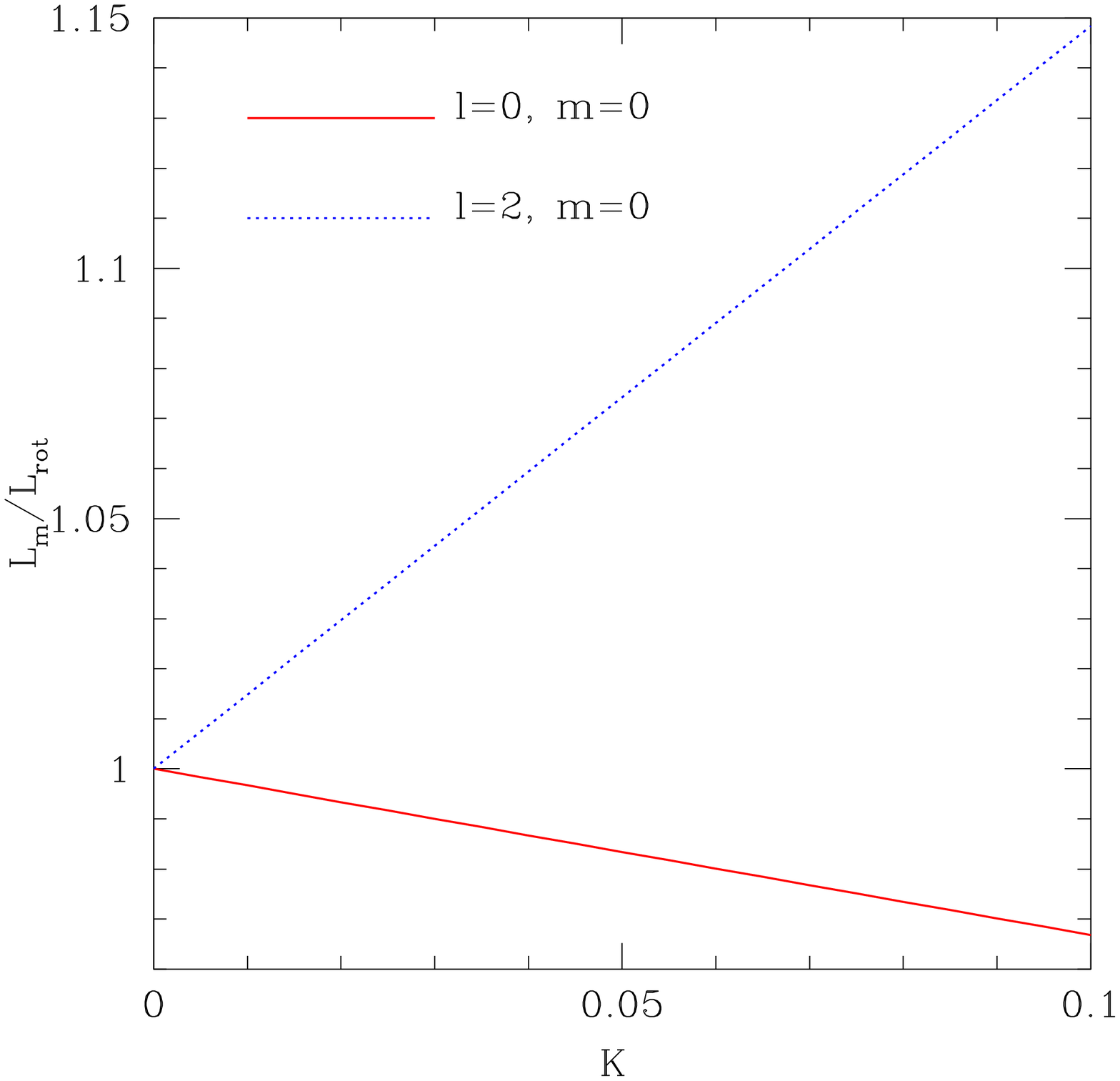}
\caption{ Left Panel: The ratio $L_{m\neq0}/L_{rot}$ as a function
of parameter $K=\tilde{\eta}(1)/\Omega R$ for modes
$(1,1)$ (continuous red
line) and $(2,1)$ (dotted blue line). Right Panel: The ratio
$L_{m=0}/L_{rot}$ as a function of parameter
$K=\tilde{\eta}(1)/\Omega R$ for modes $(0,0)$
(continuous red line) and
$(2,0)$ (dotted blue line).}
\label{fig4}
\end{center}
\end{figure}
%

%------------------------------------------
\section{Connection to the phenomenology of part-time
pulsars}
\label{Connection_to_the_phenomenology_of_part-time_pulsars}
Recently, \citet{Rea2008a} and \citet{Lyne2009} reported that a
previously known pulsar, PSR B1931+24, with a spin period of $813
\ {\rm ms}$ at the relatively large distance of $\sim 4.6 {\rm
kpc}$, when monitored long enough, shows
an intermittent radio emission,
consisting of an
active ON state, lasting for $5 - 10\ {\rm d}$,
followed by a sharp transition (happening in
less than $10 \ {\rm s}$)
to an OFF state during which the
pulsar remains undetectable for $25 - 35\ {\rm
d}$. More interestingly, the spin-down rate during the
ON state is $\dot{\nu}_{\rm ON}
= -16.3(4)\times 10^{-15} {\rm Hz\ s^{-1}}$, while, when
measured over longer periods, the average value of the
spin down is sensibly smaller. This is compatible with a
picture in which the spin down rate during the OFF state
is different from that in the ON state, and, in
particular, it is a $50 \%$ factor smaller, namely
$\dot{\nu}_{\rm OFF} =
-10.8(2)\times 10^{-15} {\rm Hz\ s^{-1}}$.
A further search for similar objects
from the Parkes Multi-Beam Survey data revealed at least four additional
objects presenting properties similar to those of PSR
B1931+24 \citep{Becker2009}, like, for instance,
PSR J1832+0031 with an ON state of $\sim 300 \ {\rm d}$
and an OFF
state of $\sim 700 \ {\rm d}$. Understanding the physical mechanism
responsible for such a remarkable phenomenology as well as
the relationship between intermittent (or ``part-time'') pulsars and
conventional radio pulsars is of course of great
interest, especially as it may
help clarifying aspects that are still obscure about
pulsar radio emission.

All of the (few) models presented so far
for explaining the phenomenology of intermittent radio pulsars
are based on the common idea that intermittent
pulsars are isolated neutron stars,
similar to conventional radio pulsars.
One of the first ideas was that
this effect could be
similar to {\it nulling}, already reported several
years ago by \cite{Backer1970}.
However, the nulling phenomenon
only lasts for a few pulse
periods and not on a timescales of tens of days
as detected for intermittent pulsars.
A second argument that was proposed
is that the intermittent phenomenology
could be due to precession, which is
an effect by which the pulsar undergoes a slow periodic wobble,
thus moving the beams of radio radiation out of our line
of sight. What ruled out this idea, however, is that
precession certainly cannot produce a transition from the
ON to the OFF state in less than $10 {\rm s}$.

A more convincing explanation proposed by \citet{Lyne2009}
and ~\cite{Gurevich2007} is that there is a global
failure of charge particle currents in the magnetosphere.
In particular, the changes in the radio emission would be
due to the presence or absence of a plasma whose current
flow provides the
expected extra torque on the star.
In this model, the open field lines above the magnetic pole
become depleted of charged radiating particles during the OFF
states and the rotational slow-down, $\dot{\nu}_{\rm OFF}$, is
produced by a torque dominated by magnetic dipole
radiation. On the contrary,
when the pulsar is ON, $\dot{\nu}_{\rm ON}$
is enhanced by an additional torque provided
by the outflowing plasma. In other words, during the ON state the
energy release is due to the current losses only, while during the OFF
state it is due to the magnetodipole vacuum radiation (in this case, it is
not the plasma-filled magnetosphere). In spite of its
plausibility, this idea suffers from the
lack of a physical mechanism for changing
the plasma flow in the magnetosphere in such a drastic
way.

What we propose here is indeed an alternative idea based on our
results about oscillating magnetospheres. As we have shown in
Sec.~\ref{Energy_losses}, the energy loss by the pulsar can be
significantly altered by the stellar oscillations. Therefore, it
is reasonable to assume that during the  ON state the stellar
oscillations create relativistic wind of charged particles by
virtue of the additional accelerating electric field. In a period
of about $10 {\rm d}$ stellar oscillations are damped and the OFF
period starts. The quasi-periodic glitch, whose driving mechanism
is still largely obscure, is the only plausible excitation
mechanism of oscillations for isolated pulsars and it would also
be responsible for the emergence of new ON states with a certain
periodicity. It is well known that the radio emission of pulsar is
negligible in the overall energy budget, usually constituting a
very tiny fraction of the pulsar spin-down rate, less then
$10^{-3}$. Most of the energy flux is carried away by the
relativistic pulsar wind and does not reveal itself during pulsar
emission. Therefore, switching radio emission on or off cannot
change the pulsar spin-down significantly enough to be detected by
observations. As a result, we propose that $\dot{\nu}_{\rm
  OFF}\approx\dot{\nu}_{\rm ON}$ and
that pulsar quasiperiodic glitch is the real
reason for i) the sudden increase of rotational energy at
the end of the OFF state and for ii)
the excitation of the stellar oscillations which switches pulsar
radio emission on.

As suggested by ~\cite{Zhang2007}, the transition from
the OFF to the ON state of intermittent pulsar
would correspond to the
reactivation of a dead pulsar above the 'death line' in
the $P - B$ diagram,
thus becoming
occasionally active only when the
conditions for pair production and coherent emission are
satisfied.
It is worth stressing that
it is very difficult to define an exact line in the $P -
B$ diagram for rotating neutron stars.
However, in a recent investigation~\cite{Ahmedov2009}
showed that oscillating but non rotating neutron star
remain below the death line for the majority of known
radio pulsars. That result, when combined to the findings
of the present study, suggest that oscillations in a
rotating star could be the key ingredient for explaining
the transient re-activation of intermittent pulsars.

The connection that we have proposed here between the
intermittent pulsar phenomenology and the presence of
oscillations in the magnetosphere of rotating magnetized
neutron stars will be exhaustively investigated in a more
quantitative way in a forthcoming paper.

%-------------------------------------------
\section{Conclusions}
\label{concl}

In this paper we have studied the astrophysical processes in the
polar cap region of the magnetosphere of an oscillating neutron
star. The background spacetime is given by the metric of
~\citet{Hartle1968} within the slow rotation approximation. The
novelties of our analysis consists in quantifying the
contributions of stellar oscillations in a general relativistic
framework. In particular, we have computed the
general-relativistic corrections to the Goldreich-Julian charge
density, to the electrostatic scalar potential and to the
component of the electric field parallel to the magnetic field
lines in the polar cap region when toroidal stellar oscillations
are present. As already remarked by~\citet{Timokhin2007}, the
effective electric charge density \ie the difference between the
Goldreich-Julian charge density $\rho_{\rm{GJ}}$ (proportional
to $\vec{\Omega}\cdot \vec{B}$ for rotating stars
and to $\vec{\omega}\cdot \vec{B}$ for oscillating stars
in the flat space-time case) and the
electric charge density (proportional to $\vec{B}$) in the
oscillating star magnetosphere is responsible for the generation
of an electric field parallel to the magnetic field lines. Such
difference vanishes only at the surface of the star while in
general it becomes significantly large at some distance $r$ from
the surface, due to the fact that $\rho$ can not compensate
$\rho_{\rm{GJ}}$. As already pointed out by~\citet{Muslimov1992},
general relativistic terms arising from the dragging of inertial
frames give very important additional contribution to this
difference. These terms depend on the radial distance from the
star as $1/r^3$ and have important influence on the value of
accelerating electric field generated in the magnetosphere near
the surface of the neutron star.

Our solutions for the accelerating electric field for the
oscillating and rotating magnetized neutron stars may have some
significant implications for pulsar polar cap models. These models
assume that charged particles are accelerated above the polar
caps, initiating pair cascades through one-photon pair creation of
photons.  The electric field induced by the stellar oscillations
becomes therefore very important. Thus, the potential drop at the
pair formation front, and the total energy gained by particles in
the open field region is larger for the oscillating star.
Since the contribution from the stellar oscillations to the
electric field depends on the amplitude of
stellar oscillations, pulsars having larger
$K=\tilde{\eta}(1)/\Omega R$
will have larger accelerating potential drops.
Our main conclusions about
oscillating and rotating neutron stars can be summarized
as follows:

\begin{enumerate}

\item
The oscillation regime of particle ejection from the stellar
surface increases the total power carried away by relativistic
primary particles relative to the purely rotating regime.
Moreover, the
fluctuation of the charge density of particles ejected from the
stellar surface modulates the particle energy along a field line.

\item The energy losses along the open magnetic field lines
in the polar cap region
and due to toroidal oscillations are
significantly larger than the rotational energy losses
for the $m=1$ modes of oscillation. In particular, the
energy losses of the
mode $(l,m)=(2,1)$ can be a factor $8$
larger than the rotational energy losses, even for an
oscillation amplitude at the star surface as small as
$\eta=0.05 \ \Omega \ R$.

\item
The oscillation-induced inhomogeneity of the physical conditions at the stellar
surface may substantially affect the global electrodynamics within
the inner magnetosphere of a neutron star.

\item
The new dependence obtained for the energy
losses on the oscillating behavior reflects in a new
relation, namely Eq.~\eqref{PP},
between the product $P\dot{P}$ and the amplitude of the
oscillation at the star surface. In cases when the moment
of inertia of the star is known with good accuracy, such
a relation will allow to fully appreciate the effects of
oscillations on pulsar magnetospheres.

\end{enumerate}

Finally, we have
proposed a connection between the phenomenology of
intermittent pulsars, characterized by the periodic
transition from active to dead periods of radio emission
in few observed sources, with the presence of an oscillating
magnetosphere. In particular, we propose
that, during the active state,
star oscillations induced by periodic glitches of the neutron star
create relativistic wind
of charged particles by virtue of the
additional accelerating electric field.
After a timescale of the order of tens of days
stellar oscillations
are damped, and the pulsar shifts below the death line in
the  $P - B$ diagram, thus entering the OFF invisible
state of intermittent pulsars.
This seminal idea, proposed here on a qualitative level,
will be further explored in a future work.

%%%%%%%%%%%%%%%%%%%%%%%%%%%%%%%%%%%%%%%%%%%%%%%%%%%%%%%%%%%%%%%%%%%%%%%%%
\section*{Acknowledgments}
%%%%%%%%%%%%%%%%%%%%%%%%%%%%%%%%%%%%%%%%%%%%%%%%%%%%%%%%%%%%%%%%%%%%%%%%%

We wish to thank Luciano Rezzolla for bringing this problem to our
attention and for many helpful discussions and comments.
This research is supported in part by Projects No. FA-F2-079 and
No. FA-F2-F061 of the UzAS. BJA acknowledges the partial financial
support from the German Academic Exchange Service
DAAD and the IAU C46-PG-EA programs. This work was supported in
part by the DFG grant SFB/Transregio~7.

\bibliographystyle{mn2e}
%\bibliography{aeireferences}

\bsp

\label{lastpage}

\end{document}